\documentclass[letterpaper]{article} 
\usepackage{aaai23}  
\usepackage{times}  
\usepackage{helvet}  
\usepackage{courier}  
\usepackage[hyphens]{url}  
\usepackage{graphicx} 
\usepackage{natbib}  
\usepackage{caption} 
\usepackage{algorithm}
\usepackage{algorithmic}
\usepackage{amsthm,amsmath,amssymb}
\usepackage{cite}
\usepackage{booktabs}
\usepackage{multirow}
\usepackage{subfigure}
\usepackage[misc]{ifsym}
\usepackage{xcolor}

\usepackage{newfloat}
\usepackage{listings}
\DeclareCaptionStyle{ruled}{labelfont=normalfont,labelsep=colon,strut=off} 
\floatstyle{ruled}
\newfloat{listing}{tb}{lst}{}
\floatname{listing}{Listing}
\title{Uniform Sequence Better: Time Interval Aware Data Augmentation\\for Sequential Recommendation}
\author {
    Yizhou Dang\textsuperscript{\rm 1},
    Enneng Yang \textsuperscript{\rm 1},
    Guibing Guo \textsuperscript{\rm 1}\thanks{Corresponding Author.},
    Linying	Jiang \textsuperscript{\rm 1},
    Xingwei Wang \textsuperscript{\rm 2}$\textsuperscript{\rm *}$\\
    Xiaoxiao Xu \textsuperscript{\rm 3},
    Qinghui Sun \textsuperscript{\rm 3},
    Hong Liu \textsuperscript{\rm 3}
}
\affiliations {
    \textsuperscript{\rm 1} Software College, Northeastern University, China \\
    \textsuperscript{\rm 2} School of Computer Science and Engineering, Northeastern University, China
    \textsuperscript{\rm 3} Alibaba Group \\
    \{dangyz,ennengyang\}@stumail.neu.edu.cn, \{guogb,jiangly\}@swc.neu.edu.cn \\ wangxw@mail.neu.edu.cn, \{xiaoxiao.xuxx,yuyang.sqh,liuhong.liu\}@alibaba-inc.com
}

\begin{document}

\maketitle

\begin{abstract}
Sequential recommendation is an important task to predict the next-item to access based on a sequence of interacted items. Most existing works learn user preference as the transition pattern from the previous item to the next one, ignoring the time interval between these two items. However, we observe that the time interval in a sequence may vary significantly different, and thus result in the ineffectiveness of user modeling due to the issue of \emph{preference drift}. In fact, we conducted an empirical study to validate this observation, and found that a sequence with uniformly distributed time interval (denoted as uniform sequence) is more beneficial for performance improvement than that with greatly varying time interval. Therefore, we propose to augment sequence data from the perspective of time interval, which is not studied in the literature. Specifically, we design five operators (Ti-Crop, Ti-Reorder, Ti-Mask, Ti-Substitute, Ti-Insert) to transform the original non-uniform sequence to uniform sequence with the consideration of variance of time intervals. Then, we devise a control strategy to execute data augmentation on item sequences in different lengths. Finally, we implement these improvements on a state-of-the-art model CoSeRec and validate our approach on four real datasets. The experimental results show that our approach reaches significantly better performance than the other 11 competing methods. Our implementation is available: https://github.com/KingGugu/TiCoSeRec.
\end{abstract}

\section{Introduction}
Sequential recommendation has become an important research task to predict an accurate item to access at the next time step for a user based on her historical data, which is often organized as a sequence of interacted items. Most existing works attempt to learn user preference as a transition pattern from the sequential data, i.e., how the previous item can be properly transited to the next item. The ordering of interacted items matters a lot in this task~\cite{2021BiCAT}. Typical networks such as Transformer are applied to model sequential data and produce recommendations~\cite{2018SASRec, 2019BERT4Rec, 2020Time, 2020Disentangled, 2021BiCAT}. 

\begin{figure}[!t]
	\centering
	\includegraphics[scale=0.34]{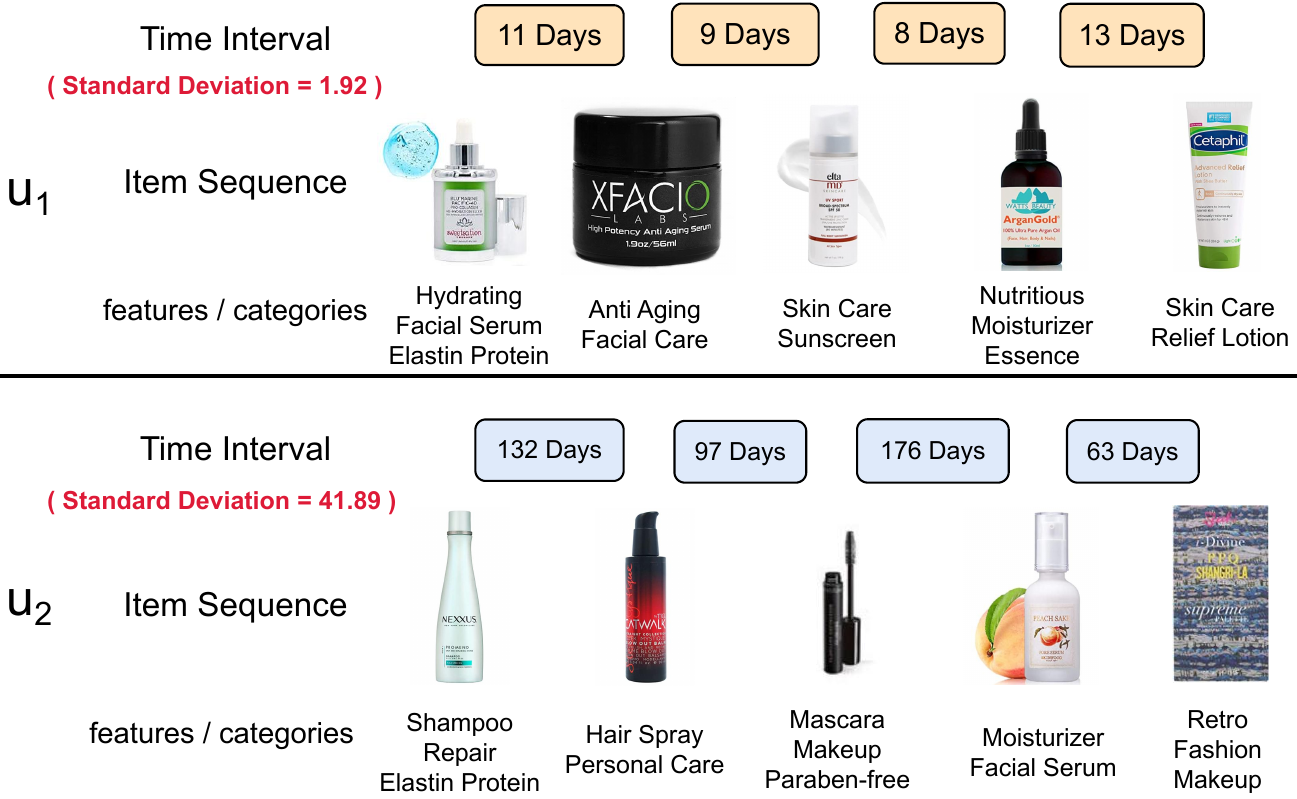}
	\caption{Two sequences drawn from the Amazon Beauty dataset with different distributions of time intervals.}
	\label{fig:example}
	\vspace{-2mm}
\end{figure}

However, the time interval between two consecutive items (in a sequence) does not receive much attention in the literature, and we believe this factor plays an important role in learning user preferences, especially when considering the issue of \emph{preference drift}, i.e., user preference may drift away as time goes by. As a matter of fact, we observe that the time interval in a sequence may vary significantly different, which may lead to the ineffectiveness of user modeling merely based on the ordering of interacted items. Figure~\ref{fig:example} shows a typical example that we drew from a real dataset (Amazon Beauty). Both users have a sequence of interacted items with the same length, but the time intervals for user $u_1$ are distributed more uniformly than those for user $u_2$. Most products purchased by user $u_1$ lie in the category of \emph{Skin} / \emph{Facial Care}, while user $u_2$ begins to purchase products in different features/categories every after a large time interval (e.g., from \emph{Hair Spray} to \emph{Mascara}, from \emph{Mascara} to \emph{Moisturizer}). It is intuitively believed that user preference can be better learned for user $u_1$ since her sequence is more uniformly distributed than the other user. We denote such a kind of sequence for user $u_1$ as `uniform sequence' and that for user $u_2$ as `non-uniform sequence'. 

In this paper, we conduct an empirical study to further validate this intuition. Specifically, we design two strategies to split the original sequential data into uniform and non-uniform sequences, and then compare the performance of four competing recommenders under different settings. The experimental results show that performance can be significantly improved with uniform sequences rather than non-uniform sequences. However, it is a commonplace that real sequences may not be uniformly distributed in terms of the time interval. To this end, we propose five operators of data augmentation (denoted as Ti-Crop, Ti-Reorder, Ti-Mask, Ti-Substitute, and Ti-Insert) to transform non-uniform original sequences to uniform sequences with the consideration of standard deviation of time intervals. Besides, we apply contrastive learning to ensure that augmented sequences preserve a high similarity with the original ones. Finally, we implement these improvements on a state-of-the-art model CoSeRec~\cite{CoSeRec} and thus form a time interval aware recommendation model called TiCoSeRec.  

To sum up, the main contributions of this paper can be presented as follows:
\begin{itemize}
	\item We conducted an empirical study to verify the idea that uniform sequences are more beneficial for performance improvement than non-uniform sequences. To the authors' best knowledge, we are the first to study the distribution of time intervals for sequential recommendation.

	\item We propose five operators of data augmentation to transform non-uniform sequences to uniform ones, and generate new sequences of high similarity to the original ones. Besides, we also devise a control strategy to deal with data augmentation for short sequences. 

	\item We conducted extensive experiments on four real datasets and demonstrated that our approach TiCoSeRec can achieve a significant performance improvement in comparison with other 11 competing models. 
\end{itemize}

\section{Related Work}
\subsection{Sequential Recommendation Models}
The existing works of sequential recommendation focus on how to effectively model the sequence of interacted items. An early solution is to treat the item sequence as a Markov Chain~\cite{MC1, 2010Factorizing}, where the next item to predict is closely related to the latest few interacted items. The limitation lies in the inability to learn the dependency in a relatively large time step. Hence, a better solution based on recurrent neural networks (RNNs) emerges since it is able to capture long- and short-term preferences via neural memory operations \cite{2015Session, 2016context, 2022cmnrec}. However, RNN cannot be well paralleled and thus will reach its performance bottleneck when dealing with a large volume of sequential data. More recently, transformer-based approaches have attracted much attention because of their convenience for parallelization and carefully designed architecture of self-attention \cite{2018SASRec, 2019BERT4Rec, 2020Time,  2020exploiting, 2021BiCAT}. However, all the works focus on the ordering sequence of items, but none of them has considered the factor of time interval between two consecutive items in a sequence, which plays an important role in preference learning, as demonstrated in the next section. Our work aims to fill up this research gap by considering the distribution of time intervals of a given sequence.

The most relevant works are the following sequential models enhanced by temporal information. \cite{2015Time} captures the recurrent temporal patterns in an extensive collection of user-item consumption pairs. Instead of taking temporal information as model input, \cite{time2} analyzes the temporal patterns of user behaviors in terms of absolute time and relative time. Later, inspired by the architecture of Transformer \cite{2017Attention}, \cite{2020Time} designs a new model by incorporating the relative time intervals and absolute positions among items with an attention mechanism. \cite{2018modeling} proposes a statistical model based on multivariate temporal point processes that jointly model complex human behaviors. \cite{Seol_2022} learns both long- and short-term preferences with the incorporation of temporal characteristics. Besides, \cite{time3} proposes a functional feature map that embeds time into high-dimensional spaces, bridging the gap between modeling time-independent and time-dependent event sequences. All the above models attempt to enhance recommenders by regarding time as side information, whereas our approach intends to enhance the sequential data via data augmentation rather than recommenders. 

\subsection{Data Augmentation for Recommendation}
Data augmentation is an effective method to improve recommendation performance. For example, \cite{2020Future} finds that future interaction can be used to augment sequential data and improve the model training. \cite{2021Context} incorporates spatial and temporal information to augment sequences for next POI recommendation. \cite{2021BiCAT} proposes a conditional generative adversarial network to generate new data that is close to the original data, whereby more data can be used for recommendation. \cite{2021CASR} revises the sequence of user behaviors from the perspective of counterfactual reasoning, i.e., by substituting some previously purchased items with other unknown items. A common issue of these approaches is that augmented data may not preserve the same quality as the original data. 

To resolve this issue, many researchers leverage self-supervised techniques (e.g., contrastive learning) to impose similarity contrast between augmented data and the original. For example, \cite{2020CL4SRec} comes up with three data augmentation operators (Crop, Reorder, Mask) for pre-training recommendation tasks, which are equipped with a contrastive learning framework. \cite{2020Self} proposes a two-stage augmentation strategy that first masks item embedding layers and then drops categorical features other than contrastive learning. \cite{2020S3Rec} also applies contrastive learning to maximize the mutual information over attributes and sequential augmentation by random masking attributes and items. \cite{2021DuoRec} contains a contrastive regularization with both the dropout-based model-level augmentation and the supervised positive sampling to construct contrastive samples. \cite{CoSeRec} integrates item similarity information for data augmentation (Insert, Substitute) with a contrastive learning objective, aiming to maximize the agreement of augmented sequences. Our work follows this line of data augmentation with contrastive regularization, but differs in that our data augmentation operators are based on time interval, which has not been studied yet. 

\section{Assumption Validation}
In this section, we will conduct an empirical study to verify our assumption that recommenders can get better performance with uniform sequences than non-uniform ones.

\ \\ \noindent \textbf{Problem Statement.}
Suppose we have user and item sets denoted by symbols $\mathcal{U}$ and $\mathcal{V}$, respectively. Each user $u \in \mathcal{U}$ is associated with a sequence of interacted items in chronological order $S_u = (v_1,\ldots,v_j,\ldots,v_ {N})$, where $v_j \in \mathcal{V}$ indicate the item that user $u$ has interacted with at time step $j$ and $N$ is the sequence length. Each item sequence $S_u$ is associated with a sequence of time intervals $T_u = (t_1,\ldots,t_j,\ldots,t_ {N-1})$, where $t_j$ is the time interval between the timestamps of items $v_j$ and $v_{j+1}$. In this paper, we define \textbf{uniform (non-uniform) sequence} as the sequence of items $S_u$ where its corresponding sequence of time intervals $T_u$ has a relatively small (large) value of standard deviation. That is, the item sequence $S_u$ has a uniformly (non-uniformly) distributed sequence of time intervals $T_u$. 

Therefore, the sequential recommendation task in this paper can be formulated as follows. Given the sequences of interacted items $S_u$ and time intervals $T_u$, how to accurately predict the most possible item $v$ that user $u$ will interact with at time step $N$ + 1, given by:  
\begin{equation}
	\underset{v \in \mathcal{V}}{\arg \max}\; P\left(v_{N+1}=v \mid S_{u}, T_u\right)
\end{equation}

\noindent \textbf{Datasets.}
We adopt four public datasets for our experimentations throughout this paper, including three Amazon product rating 5-core datasets (Beauty, Sports, Home) where each user or item has at least five ratings~\cite{Amazon}, and a Yelp review dataset~\footnote{https://www.yelp.com/dataset}. 
To speed-up experimentation for Yelp, we preserve the sequence data recorded in 2019 and filter out the users or items with less than five ratings, i.e., the same as Amazon datasets. The statistics of our processed datasets are given in Table~\ref{tab:datasets}. Leave-one-out strategy is used to split datasets. That is, we preserve the last and second last records for each user as the testing and validation data, while the rest is taken as the training data.

\begin{table}[!htbp]
	\centering
	\scalebox{0.72}{
		\begin{tabular}{c|rrrr}
			\toprule \textbf{{Dataset}} & \textbf{Yelp} &\textbf{Beauty} & \textbf{Sports} & \textbf{Home} \\
			\midrule 
			{Users} & 19,855 & 22,363 & 35,598 & 66,370 \\
			 {Items} & 14,541 & 12,101 & 18,357 & 28,237 \\
			 {Interactions} & 207,045 & 198,502 & 296,337 & 551,682 \\
			 {Sparsity} & 99.92\% & 99.92\% & 99.95\% & 99.97\% \\
			 {Average Sequence Length} & 10.42 & 8.87 & 8.32 & 8.29 \\
			\bottomrule
	\end{tabular}}
	\caption{The statistics of our datasets.}
	\label{tab:datasets}
\end{table}

\begin{figure}[!t]
	\centering
	\includegraphics[scale=0.40]{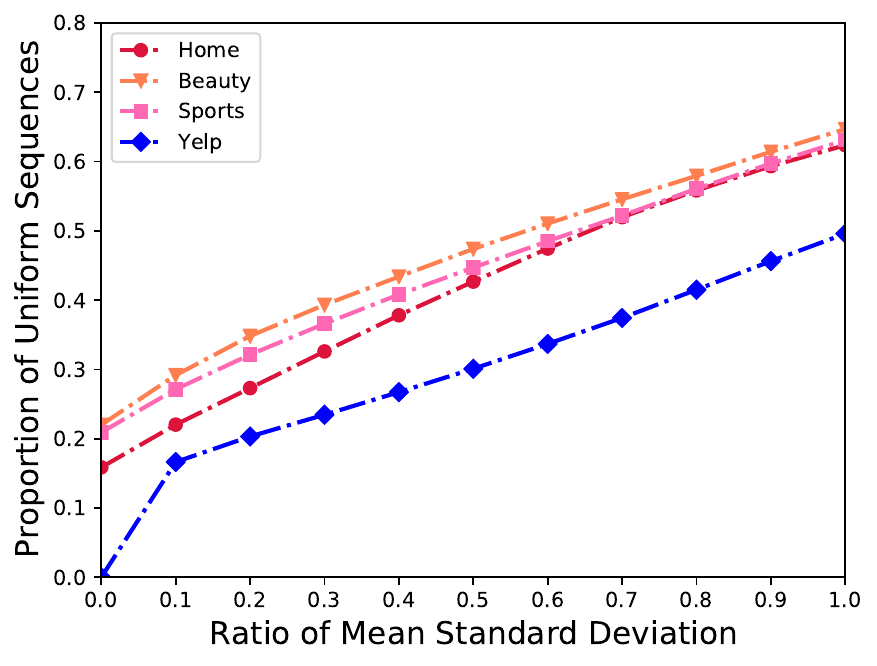}
	\caption{Proportion of uniform sequences in four datasets.}
	\label{fig:ratio}
 \vspace{-2mm}
\end{figure}

\noindent \textbf{Dataset Analysis.} We determine a given sequence as a uniform sequence if its standard deviation is smaller than a threshold. We set the threshold as the \emph{ratio} of mean standard deviation over all users. Then, we can calculate the percentage of uniform sequences over all the data, and we repeat this procedure by adjusting different ratio values. The results are illustrated in Figure~\ref{fig:ratio}. 
When the threshold increases linearly, the ratio growth of uniform sequences slows down, indicating that the residual non-uniform sequences have large standard deviation. When we set the ratio as 0, Amazon datasets have around 20\% sequences that have the same standard deviation (as 0). This is because those users have very short sequences and all the sequence items are purchased at the same time. When the ratio is set to 1, the percentage of uniform sequences will reach the maximum value. Specifically, over 60\% sequences can be regarded as uniform in Amazon datasets, while around 50\% sequences as uniform in Yelp dataset. In other words, non-uniform sequences occupy a large portion of the overall data (40-50\%).

\ \\ \noindent \textbf{Dataset Partition.} To compare the performance with uniform and non-uniform sequences, we have to partition the original dataset into two subsets, one with uniform sequences (the subset denoted by $\mathbb{U}$) and the other with non-uniform ones (the subset denoted by $\mathbb{N}$). For a fair comparison, we need to make sure that the two subsets have a similar amount of data. To start with, we rank all the item sequences according to the standard variance of time intervals. The smaller the variance, the higher the ranking. We design three partition strategies to split our datasets. 
\begin{itemize}
	\item \textbf{Strategy S: subsets $\mathbb{U}$ and $\mathbb{N}$ have equal size of item sequences.} Specifically, subset $\mathbb{U}$ consists of item sequences that rank at the top 50\% of all the item sequences, while subset $\mathbb{N}$ contains the rest sequences. Since the length of item sequences is different from one to other, these two subsets may not have the same size of user-item interactions.
	\item \textbf{Strategy I: subsets $\mathbb{U}$ and $\mathbb{N}$ have equal size of user-item interactions.} Specifically, we add the top-ranked item sequences to subset $\mathbb{U}$ until half of all user-item interactions are involved, and the rest is added to subset $\mathbb{N}$. 
	\item \textbf{Strategy Random: randomly sample half of all the item sequences.} Specifically, item sequence is taken selected into the subset by random sampling, and we do not consider the ranking of item sequences. 
\end{itemize}
Random strategy is used as a baseline for better comparison. That is, let the three strategies perform model training under the similar amount of data.

\ \\ \noindent \textbf{Models and Evaluation Metrics.}
Four classic sequential recommendation models are adopted for empirical study including \textbf{GRU4Rec} \cite{2015GRU}, \textbf{SASRec} \cite{2018SASRec}, \textbf{TiSASRec} \cite{2020Time} and \textbf{CoSeRec} \cite{CoSeRec}.
We evaluate the performance by NDCG@N and Hit@N, where N is set to 10. We leave the parameter settings to the later section. 

\begin{table}[!tb]
  \centering
\scalebox{0.55}{
    \begin{tabular}{c|c|cccccccc}
    \toprule
    \multicolumn{1}{c|}{\multirow{2}[2]{*}{Dataset}} & \multirow{2}[2]{*}{Strategy} & \multicolumn{2}{c}{\textbf{GRU4Rec}} & \multicolumn{2}{c}{\textbf{SASRec}} & \multicolumn{2}{c}{\textbf{TiSASRec}} & \multicolumn{2}{c}{\textbf{CoSeRec}} \\
      & \multicolumn{1}{c|}{} & \multicolumn{1}{c}{NDCG} & \multicolumn{1}{c}{Hit} & \multicolumn{1}{c}{NDCG} & \multicolumn{1}{c}{Hit} & \multicolumn{1}{c}{NDCG} & \multicolumn{1}{c}{Hit} & \multicolumn{1}{c}{NDCG} & \multicolumn{1}{c}{Hit} \\
    \midrule
    \multirow{7}[6]{*}{Beauty} & random & 0.0217 & 0.0455 & 0.0291 & 0.0599  & 0.0295 & 0.0552  & 0.0317 & 0.0575 \\
\cmidrule{2-10} & S: $\mathbb{U}$ & \textbf{0.0243} & \textbf{0.0514} & \textbf{0.0315} & \textbf{0.0632} & \textbf{0.0325} & \textbf{0.0603} & \textbf{0.0352} & \textbf{0.0674} \\
      & S: $\mathbb{N}$ & 0.0202  & 0.0421  & 0.0269  & 0.0584  & 0.0279  & 0.0523  & 0.0292  & 0.0523  \\
      & improve & 20.30\% & 22.09\% & 17.10\% & 8.22\% & 16.49\% & 15.30\% & 20.55\% & 28.87\% \\
\cmidrule{2-10} & I: $\mathbb{U}$ & \textbf{0.0231} & \textbf{0.0498} & \textbf{0.0311} & \textbf{0.0629} & \textbf{0.0337} & \textbf{0.0621} & \textbf{0.0365} & \textbf{0.0642} \\
      & I: $\mathbb{N}$ & 0.0183  & 0.0398  & 0.0246  & 0.0543  & 0.0266  & 0.0502  & 0.0284  & 0.0533  \\
      & improve & 26.23\% & 25.13\% & 26.42\% & 15.84\% & 26.69\% & 23.71\% & 28.52\% & 20.45\% \\
    \midrule
    \multirow{7}[6]{*}{Sports} & random & 0.0156 & 0.0305 & 0.0159 & 0.0325 & 0.0173 & 0.0297  & 0.0185  & 0.0324  \\
\cmidrule{2-10} & S: $\mathbb{U}$ & \textbf{0.0169} & \textbf{0.0335} & \textbf{0.0165} & \textbf{0.0351} & \textbf{0.0186} & \textbf{0.0326} & \textbf{0.0203} & \textbf{0.0364} \\
      & S: $\mathbb{N}$ & 0.0147  & 0.0272  & 0.0151  & 0.0301  & 0.0152  & 0.0263  & 0.0168  & 0.0297  \\
      & improve & 14.97\% & 23.16\% & 9.27\% & 16.61\% & 22.37\% & 23.95\% & 20.83\% & 22.56\% \\
\cmidrule{2-10} & I: $\mathbb{U}$ & \textbf{0.0168} & \textbf{0.0338} & \textbf{0.0171} & \textbf{0.0358} & \textbf{0.0192} & \textbf{0.0338} & \textbf{0.0196} & \textbf{0.0349} \\
      & I: $\mathbb{N}$ & 0.0142  & 0.0269  & 0.0148  & 0.0311  & 0.0145  & 0.0266  & 0.0176  & 0.0288  \\
      & improve & 18.31\% & 25.65\% & 15.54\% & 15.11\% & 32.41\% & 27.07\% & 11.36\% & 21.18\% \\
    \midrule
    \multirow{7}[6]{*}{Home} & random & 0.0098 & 0.0140  & 0.0112 & 0.0164 & 0.0086 & 0.0133  & 0.00102 & 0.0175 \\
\cmidrule{2-10} & S: $\mathbb{U}$ & \textbf{0.0111} & \textbf{0.0159} & \textbf{0.0125} & \textbf{0.0179} & \textbf{0.0096} & \textbf{0.0146} & \textbf{0.0124} & \textbf{0.0183} \\
      & S: $\mathbb{N}$ & 0.0087  & 0.0133  & 0.0108  & 0.0155 & 0.0081  & 0.0122  & 0.0099  & 0.0152  \\
      & improve & 27.59\% & 19.55\% & 15.74\% & 15.48\% & 18.52\% & 19.67\% & 25.25\% & 20.39\% \\
\cmidrule{2-10} & I: $\mathbb{U}$ & \textbf{0.0109} & \textbf{0.0146} & \textbf{0.0131} & \textbf{0.0180} & \textbf{0.0101} & \textbf{0.0144} & \textbf{0.0111} & \textbf{0.0189} \\
      & I: $\mathbb{N}$ & 0.0091  & 0.0129  & 0.0105  & 0.0147  & 0.0079  & 0.0118  & 0.0096  & 0.0161  \\
      & improve & 19.78\% & 13.18\% & 24.76\% & 22.45\% & 27.85\% & 22.03\% & 15.63\% & 17.39\% \\
    \midrule
    \multirow{7}[6]{*}{Yelp} & random & 0.0233 & 0.0463 & 0.0296 & 0.0523 & 0.0281 & 0.0489  & 0.0256  & 0.0503 \\
\cmidrule{2-10} & S: $\mathbb{U}$ & \textbf{0.0251} & \textbf{0.0498} & \textbf{0.0334} & \textbf{0.0569} & \textbf{0.0296} & \textbf{0.0527} & \textbf{0.0283} & \textbf{0.0568} \\
      & S: $\mathbb{N}$ & 0.0198  & 0.0437  & 0.0276  & 0.0489  & 0.0258  & 0.0447  & 0.0241  & 0.0475  \\
      & improve & 26.77\% & 13.96\% & 21.01\% & 16.36\% & 14.73\% & 17.90\% & 17.43\% & 19.58\% \\
\cmidrule{2-10} & I: $\mathbb{U}$ & \textbf{0.0243} & \textbf{0.0512} & \textbf{0.0325} & \textbf{0.0578} & \textbf{0.0278} & \textbf{0.0506} & \textbf{0.0278} & \textbf{0.0574} \\
      & I: $\mathbb{N}$ & 0.0204  & 0.0420  & 0.0270  & 0.0476  & 0.0254  & 0.0441  & 0.0247  & 0.0466  \\
      & improve & 19.12\% & 21.90\% & 20.37\% & 21.43\% & 9.45\% & 14.74\% & 12.55\% & 23.18\% \\
    \bottomrule
    \end{tabular}}%
    \caption{Performance of sequential models with the randomly sampled (`random'), uniform (`$\mathbb{U}$') and non-uniform (`$\mathbb{N}$') subsets across different sampling strategies (`S', `I').}
  \label{tab:empirical}%
  \vspace{-5mm}
\end{table}%

\ \\ \noindent \textbf{Analysis and Discussion.} Table~\ref{tab:empirical} depicts all the recommendation performance on randomly sampled half of original datasets (denoted by `random') and two subsets($\mathbb{U},\mathbb{N}$) across two partition strategies(`S', `I'). For easy discussion, we denote different settings by the label `Strategy: Subset'. For example, row `S: $\mathbb{U}$' indicates the performance obtained on the subset $\mathbb{U}$ under strategy S. Besides, row `improve' computes the percentage of improvements achieved by `S(I): $\mathbb{U}$' relative to `S(I): $\mathbb{N}$'. 

We can find a few observations from Table~\ref{tab:empirical}.Firstly, the performance with uniform subsets ($\mathbb{U}$) is generally better than that with non-uniform subsets ($\mathbb{N}$) under both strategies (S, I) and across all the datasets. Specifically, there are 54 cases (out of 64 pairs of comparison) with improvements greater than 15\%, and 33 cases (around 51.6\% of all cases) with improvements greater than 20\%, and the maximum improvement goes up to 32.41\%. There are only 3 cases (out of 64) lower than 10\%, and the minimum improvement is 8.22\%. Secondly, the performance of subset `random' lies in the middle of subsets $\mathbb{U}$ and $\mathbb{N}$ under both strategies S and I. This is because subset `random' can be regarded as a random mixture of subsets $\mathbb{U}$ and $\mathbb{N}$. Lastly, the best selection between strategies S and I is dataset specific. Specifically, strategy I gets greater percentage of performance improvements than strategy S on three datasets (Beauty, Sports, Yelp), but gets worse results on the rest dataset (Home).

\ \\ \noindent \textbf{Summary.} In real datasets, there are many item sequences with a large variance in terms of time intervals, and the percentage is up to 40-50\% even at a loose threshold. In our later experimentation, we will show that only top 20-30\% ranked item sequences should be regarded as uniform while the other sequences need to be augmented for better performance. Our basic assumption is well validated since the subset with uniform sequences shows better performance than that with non-uniform sequences. Hence, it is necessary and valuable to transform non-uniform sequences to uniform ones for better performance via data augmentation.

\begin{figure*}[!tb]
	\centering
	\includegraphics[scale=0.28]{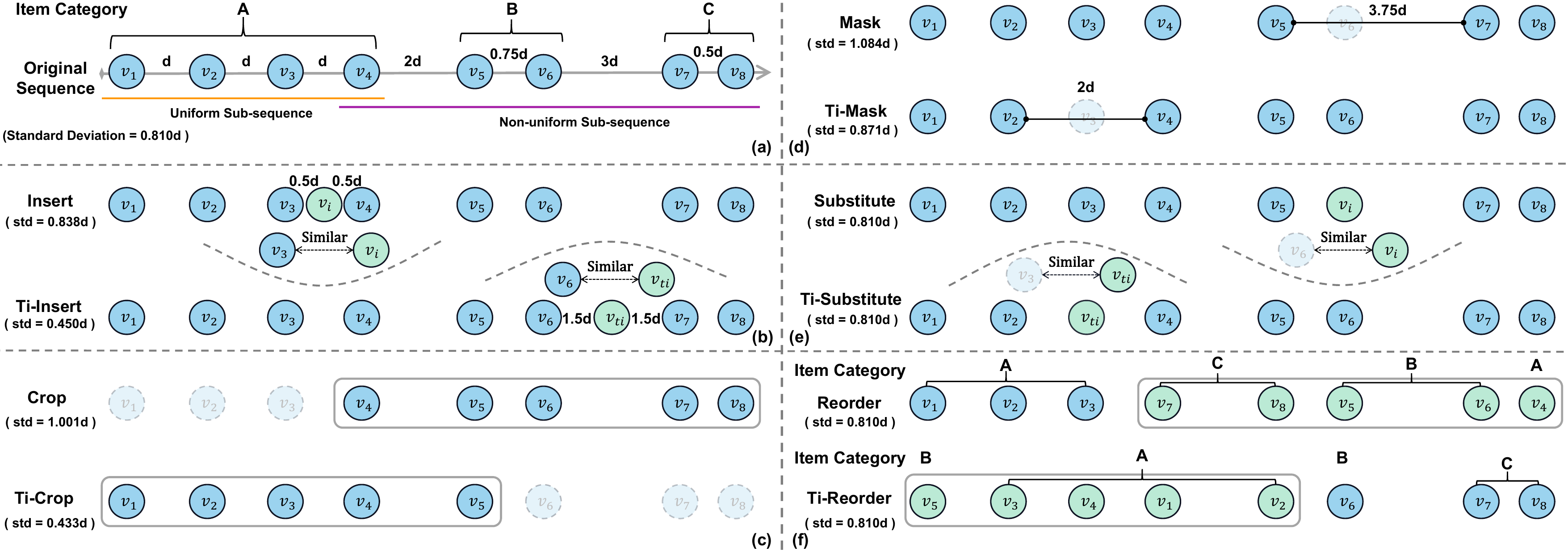}
	\caption{An illustration of the traditional and proposed data operators to augment sequential data by time intervals in unit $d$, where green nodes represent the fake items either not interacted with or interacted but at a fake time step.}
	\label{fig:aug}
\end{figure*} 

\section{Ours: Data Augmentation by Time Intervals}
As we previously discussed, we generally label the top $\sigma$ ratio of ranked item sequences as uniform, and the rest as non-uniform sequences that need to be augmented for better performance. Assume we have $m$ item sequences in total, then there are $m(1-\sigma)$ item sequences to be selected as non-uniform sequences. 

For data augmentation, there are five widely used operators in the literature, including Crop, Mask, Insert, Reorder and Substitute. These operators perform data augmentation and generate new sequences by randomly selecting target items to modify. However, such a process cannot guarantee that a newly generated sequence will possess a uniform distribution of time intervals, i.e., becoming a uniform sequence. Therefore, in this section, we will introduce five new data operators with time intervals (Ti-Crop, Ti-Mask, Ti-Insert, Ti-Reorder, Ti-Substitute) that aim to generate better item sequences by carefully selecting the most proper target items to operate. A typical comparison between our operators and traditional ones is given in Figure~\ref{fig:aug}. 

\ \\ \noindent \textbf{Ti-Insert (TI).} This operator contains two steps. It first selects $k=\beta N$ target positions $\left(p_1,p_2,\ldots,p_k\right)$ for item insertion from a given sequence, where $N$ is the sequence length and $\beta \in[0, 1]$ is the insertion ratio. Then each target position will be inserted with an item that has not been interacted with by the user but is correlated with contextual items in the sequence. We adopt the same method as \cite{CoSeRec} to select the items to insert. 

As illustrated in Figure~\ref{fig:aug}(a), we have a given sequence containing two sub-sequences: one uniform and the other non-uniform. According to the time intervals among items (in unit $d$), we may classify these items into three categories: $A, B, C$. Thus, the preference pattern can be represented by the transfer of item categories, i.e., $A\rightarrow B\rightarrow C$. 

In Figure~\ref{fig:aug}(b), since the traditional approach (Insert) selects the target position randomly, the newly generated sequence may destroy the uniform sub-sequence after inserting item $v_i$, resulting in an even worse situation. In contrast, our approach (Ti-Insert) will insert a similar item $v_{ti}$ to the right position such that the non-uniform sub-sequence becomes more uniform. The underlying assumption is that item $v_{ti}$ was accessed by the user in the middle timestamp between those of items $v_6$ and $v_7$. In this way, the original time interval (denoted by $d$) between items $v_6$ and $v_7$ is substituted by the new time intervals (value $d/2$) between $v_6$, $v_7$ and $v_{ti}$ in the new sequence, leading to a more uniform distribution of time intervals. To this end, we first sort the time intervals of sequence $T_u$ from large to small (denoted by function $\mbox{descend}(T_u)$), and then preserve the top-$k$ indices as the target positions. Formally, our positions are determined by:
\begin{equation}
\left(p_1,p_2,\ldots,p_k\right)=\mbox{top-$k$-indices}\left(\mbox{descend}(T_u) \right).
\label{equation:ti-insert}
\end{equation}

\ \\ \noindent \textbf{Ti-Crop (TC).} This operator first selects a cutoff position $p$ from which a continuous series of items of length $c=\eta N$ will be preserved as a new sequence, where $\eta \in[0, 1]$ is a ratio parameter. Traditional crop operation selects position $p$ in a random manner. For example, Figure~\ref{fig:aug}(c) sets the cutoff position at item $v_4$, and the length of the new sequence is 5.
The cropped sequence obtains (std = $1.001d$), which is much smaller than the original one. To further enhance the effect of cropping, we suggest selecting the sub-sequence (from all the possible sub-sequences) that meets the requirement of length $c$ and has the minimum value of standard deviation. Specifically, we first generate a set of sub-sequences that has the length $c$ (denoted as $\mbox{sub-sequences}(S_u, c)$), and then compute the standard deviation (i.e., $\mbox{std}$) of each sub-sequence. Our Ti-Crop operation can be formulated as: \begin{equation}
    p = \underset{p\in [1,N-1]}{\arg \min } \mbox{ std}\left(\mbox{sub-sequences}(S_u, c, p)\right).
\end{equation} 
For the example in Figure~\ref{fig:aug}(c), Ti-Crop will select the most uniform sub-sequence $(v_1, v_2, v_3, v_4, v_5)$ in length 5, which has the minimum standard deviation $0.433d$. The obtained std is smaller than that of traditional Crop (i.e., $1.001d$).

\ \\ \noindent \textbf{Ti-Mask (TM).} This operator is similar to Ti-Insert as it also needs to determine a number ($h=\mu N$) of target positions to modify, but differs in that masking operation is to delete (rather than insert) items from the original sequence. Parameter $\mu\in[0,1]$ is to specify the ratio of items to mask. Specifically, we first sort the time intervals of sequence $T_u$ from small to large (denoted by function $\mbox{ascend}(T_u)$), and then preserve the top-$h$ indices as the target positions. Mathematically, our positions are determined by:
\begin{equation}
\left(p_1,p_2,\ldots,p_h\right)=\mbox{top-$h$-indices}\left(\mbox{ascend}(T_u) \right).
\label{equation:ti-mask}
\end{equation}

Item masking will increase the distance (time interval) between the items before and after the masked item, and thus change the characteristics of overall standard deviation. Take Figure~\ref{fig:aug}(d) as an example, traditional Mask selects item $v_6$ to mask, obtaining a new sequence with standard deviation increasing from $0.810d$ to $1.084d$. As a matter of fact, the distance (i.e., $3d$) between items $v_6$ and $v_7$ is the greatest one among all the time intervals. User preference may drift to some extent in this case. Masking item $v_6$ will make it harder to learn user preference as the new maximum distance increases from $3d$ to $3.75d$. The larger distances between two items, the greater possibility they differ in more aspects. Differently, Ti-Mask operates on the items with much smaller time intervals, and the standard deviation gets decreased from $0.810d$ to $0.871d$ after masking item $v_3$, i.e., becoming a more uniform sequence. 

\ \\ \noindent \textbf{Ti-Substitute (TS).} This operator have the same strategy to determine target positions as Ti-Mask. The only difference is the number of candidate positions in Eq.\eqref{equation:ti-mask}. For item selection for replacement, it adopts the same approach as used in Ti-Insert. That is, we replace the original items with fake yet correlated items. In Figure~\ref{fig:aug}(e), traditional operator has a high chance to operate on the critical items (such as items $v_6$ and $v_7$), causing a deterioration in capturing the changes of user preference. In contrast, Ti-Substitute focuses on the items with smaller time intervals, easier to preserve the similar preference pattern as the original sequence. In general, this operator will obtain new sequences by imposing the minimum changes (via substitution) to the original ones. 

\begin{table*}[!t]
	\centering
	\scalebox{0.605}{
	\begin{tabular}{c|c|cc|cccccc|ccc|c|r}
		\toprule
		\multicolumn{1}{c|}{\multirow{2}[4]{*}{\textbf{Dataset}}} & \multicolumn{1}{c|}{\multirow{2}[4]{*}{\textbf{Metric}}} & \multicolumn{2}{c|}{\textbf{Non-sequential Models}} & \multicolumn{6}{c|}{\textbf{Sequential Models without Data Augmentation}} & \multicolumn{4}{c|}{\textbf{Sequential Models with Data Augmentation}} & \multicolumn{1}{c}{\multirow{2}[4]{*}{\textbf{Improve}}} \\
		\cmidrule{3-14} 
		& & BPR & LightGCN & STAMP & GRU4Rec & BERT4Rec & TiSASRec & SASRec & LightSANs & CL4SRec & CoSeRec & DuoRec & TiCoSeRec & \\
		\midrule
		\multicolumn{1}{c|}{\multirow{4}[1]{*}{Beauty}}
		& Hit@10 & 0.0438 & 0.0540 & 0.0508 & 0.0533 & 0.0526 & 0.0552 & 0.0562 & 0.0567 & 0.0569 & 0.0675 & \underline{0.0686} & \textbf{0.0737} & 7.43\% \\
		& Hit@20 & 0.0626 & 0.0803 & 0.0743 & 0.0767 & 0.0825 & 0.0845 & 0.0817 & 0.0857 & 0.0885 & 0.1015 & \underline{0.1022} & \textbf{0.1079} & 5.58\% \\
		& NDCG@10 & 0.0229 & 0.0292 & 0.0283 & 0.0289 & 0.0263 & 0.0294 & 0.0305 & 0.0290 & 0.0277 & \underline{0.0381} & 0.0359 & \textbf{0.0421} & 10.50\% \\
		& NDCG@20 & 0.0276 & 0.0358 & 0.0342 & 0.0355 & 0.0338 & 0.0372 & 0.0369 & 0.0363 & 0.0356 & 0.0467 & \underline{0.0470} & \textbf{0.0499} & 6.17\% \\
		\midrule
		\multicolumn{1}{c|}{\multirow{4}[2]{*}{Sports}}
		& Hit@10 & 0.0291 & 0.0378 & 0.0327 & 0.0347 & 0.0332 & 0.0297 & 0.0401 & \underline{0.0437} & 0.0414 & 0.0421 & 0.0392 & \textbf{0.0507} & 16.02\% \\
		& Hit@20 & 0.0460 & 0.0578 & 0.0474 & 0.0557 & 0.0538 & 0.0469 & 0.0565 & \underline{0.0668} & 0.0637 & 0.0629 & 0.0630 & \textbf{0.0767} & 14.82\% \\
		& NDCG@10 & 0.0151 & 0.0200 & 0.0185 & 0.0185 & 0.0164 & 0.0158 & 0.0181 & 0.0209 & 0.0215 & \underline{0.0241} & 0.0195 & \textbf{0.0282} & 17.01\% \\
		& NDCG@20 & 0.0194 & 0.0251 & 0.0222 & 0.0229 & 0.0216 & 0.0201 & 0.0224 & 0.0245 & 0.0271 & \underline{0.0292} & 0.0256 & \textbf{0.0345} & 18.15\% \\
		\midrule
		\multicolumn{1}{c|}{\multirow{4}[2]{*}{Home}}
		& Hit@10 & 0.0064 & 0.0096 & 0.0178 & 0.0143 & 0.0165 & 0.0129 & 0.0191 & 0.0123 & 0.0184 & 0.0232 & \underline{0.0251} & \textbf{0.0267} & 6.46\% \\
		& Hit@20 & 0.0104 & 0.0155 & 0.0247 & 0.0227 & 0.0224 & 0.0200 & 0.0263 & 0.0187 & 0.0285 & 0.0336 & \underline{0.0347} & \textbf{0.0391} & 12.76\% \\
		& NDCG@10 & 0.0031 & 0.0051 & 0.0121 & 0.0070 & 0.0089 & 0.0066 & 0.0121 & 0.0068 & 0.0092 & \underline{0.0136} & 0.0124 & \textbf{0.0153} & 12.62\% \\
		& NDCG@20 & 0.0043 & 0.0065 & 0.0133 & 0.0090 & 0.0103 & 0.0085 & 0.0131 & 0.0085 & 0.0118 & \underline{0.0165} & 0.0152 & \textbf{0.0184} & 11.52\% \\
		\midrule
		\multirow{4}[2]{*}{Yelp}
		& Hit@10 & 0.0478 & 0.0542 & 0.0412 & 0.0486 & 0.0508 & 0.0594 & 0.0572 & 0.0624 & 0.0581 & 0.0613 & \underline{0.0627} & \textbf{0.0658} & 4.94\% \\
		& Hit@20 & 0.0683 & 0.0760 & 0.0668 & 0.0848 & 0.0786 & 0.0941 & 0.0908 & 0.0963 & 0.0896 & 0.0965 & \underline{0.0976} & \textbf{0.1032} & 5.74\% \\
		& NDCG@10 & 0.0302 & 0.0328 & 0.0214 & 0.0239 & 0.0279 & 0.0323 & 0.0311 & 0.0339 & 0.0312 & 0.0322 & \underline{0.0345} & \textbf{0.0371} & 7.54\% \\
		& NDCG@20 & 0.0354 & 0.0383 & 0.0278 & 0.0330 & 0.0352 & 0.0417 & 0.0396 & 0.0426 & 0.0390 & 0.0425 & \underline{0.0432} & \textbf{0.0459} & 6.25\% \\
		\bottomrule
	\end{tabular}}
	\caption{Performance comparisons of different methods. The best score is bolded and the second best is underlined in each row. The last column is the relative improvement compared with the best baseline result.}
	\label{tab:performance}
\end{table*}

\ \\ \noindent \textbf{Ti-Reorder (TR).} 
This operator creates a new sequence by shuffling the positions of items in a sub-sequence, which is selected from the given original sequence. The selection of such a sub-sequence is the same as Ti-Crop. The problem of random shuffling is that it may destroy the preference pattern of the original sequence. For example, in Figure~\ref{fig:aug}(f) the new category pattern of sequential items after shuffling become ($A\rightarrow C\rightarrow B \rightarrow A$), while the category pattern of accessing before shuffling is ($A\rightarrow B\rightarrow C)$ as shown in Figure~\ref{fig:aug}(a). In other words, such a straightforward reordering operation is not a success in terms of category pattern. As a comparison, since Ti-Crop will select the sub-sequence with minimum standard deviation, it has a relatively high chance to preserve similar preference pattern after data reordering. For example, we get a new category pattern in Figure~\ref{fig:aug}(f) is $B\rightarrow A\rightarrow B \rightarrow C$, which is similar to the original one. 

\ \\ \noindent \textbf{Remarks.}  From Figure~\ref{fig:aug}, we can find that three operators (TI, TC, TM) will effectively reduce the standard deviation of time intervals, i.e., transforming non-uniform sequences to uniform ones. In the contrast, the rest two operators (TS, TR) will not influence the value of standard deviation, but aim to preserve the preference pattern as close to original sequences. We explain their usefulness as the ways to stop data augmentation from generating even more non-uniform sequences. To sum up, our operators will generate valuable new sequential data by either reducing the variance of time intervals, or preserving a high-quality of preference pattern.   

\ \\ \noindent \textbf{Data Augmentation by Sequence Length.} Different non-uniform sequences may vary in sequence length. Short sequences are more sensitive to the data augmentation, and thus inappropriate for certain data operations. Inspired by \cite{CoSeRec}, we devised the following strategy for better augmentation. Specifically, we use a hyper-parameter $K$ as the threshold to determine whether a sequence is short, and then perform our data augmentation as follows:
\begin{equation}
\resizebox{0.43\textwidth}{!}{$
S_{u}' = \begin{cases}\mbox{aug}\left(S_{u}, \mbox{op} \right),  \mbox{op}\in \{\mathrm{TS}, \mathrm{TI}, \mathrm{TM}\}, & N \leq K \\ \mbox{aug}\left(S_{u},\mbox{op}\right), \mbox{op} \in \{\mathrm{TS}, \mathrm{TI}, \mathrm{TM}, \mathrm{TC}, \mathrm{TR}\}, & N > K\end{cases}$}
\end{equation}
where $\mbox{aug}(S_u, \mbox{op})$ is a function to perform a specific data operator $\mbox{op}$ on a given sequence $S_u$, and returns a new sequence $S_u'$. Note that we have also tested different combinations of data operators for short sequences, and find that the above combination (TS, TI, TM) works the best. 

\section{Experimentation}

\textbf{Comparison methods.} To evaluate the effectiveness of our data augmentation, we apply it to a state-of-the-art sequential model CoSeRec~\cite{CoSeRec} and relabel this variant as \textbf{TiCoSeRec} for ease of discussion. Then, we compare with the following three groups of recommendation models. 
\begin{itemize}
    \item \textbf{Non-sequential models} include {BPR} \cite{2009BPR} and {LightGCN} \cite{2020LightGCN}. 
    \item \textbf{Sequential models without data augmentation} include {GRU4Rec} \cite{2015GRU}, {STAMP} \cite{2018STAMP}, {SASRec} \cite{2018SASRec}, {BERT4Rec} \cite{2019BERT4Rec}, {TiSASRec} \cite{2020Time} and {LightSANs} \cite{2021LightSANs}.
    \item \textbf{Sequential models with data augmentation} include {CL4SRec} \cite{2020CL4SRec},  {CoSeRec} \cite{CoSeRec} and {DuoRec} \cite{2021DuoRec}.
\end{itemize}

 \begin{figure*}[!tb]
	\centering
	\subfigure[Effect of data operators on Sports]{
	\includegraphics[scale=0.31]{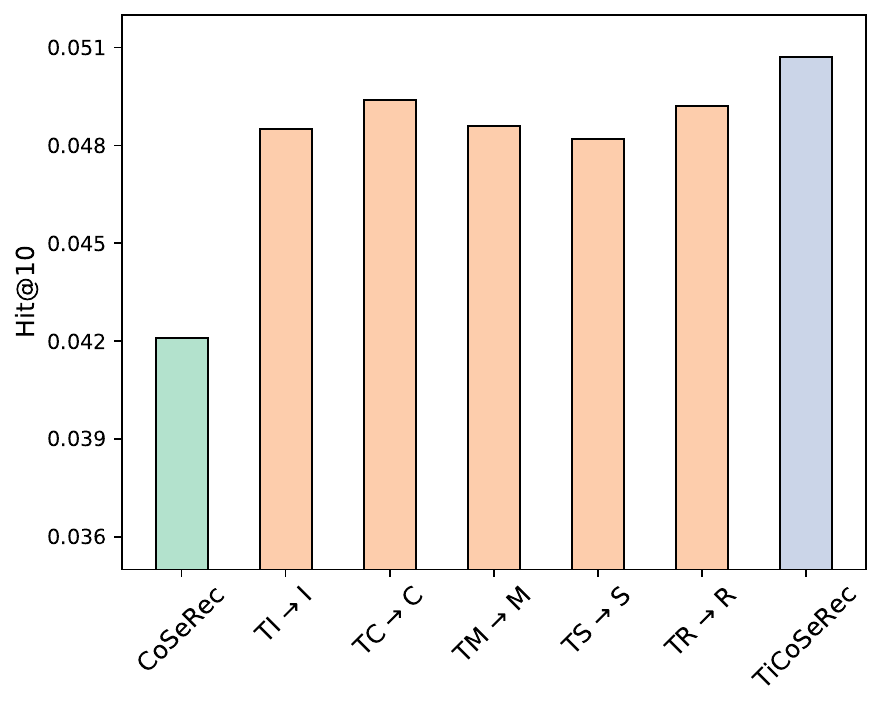}
	}
	\subfigure[Effect of data operators on Home]{
	\includegraphics[scale=0.31]{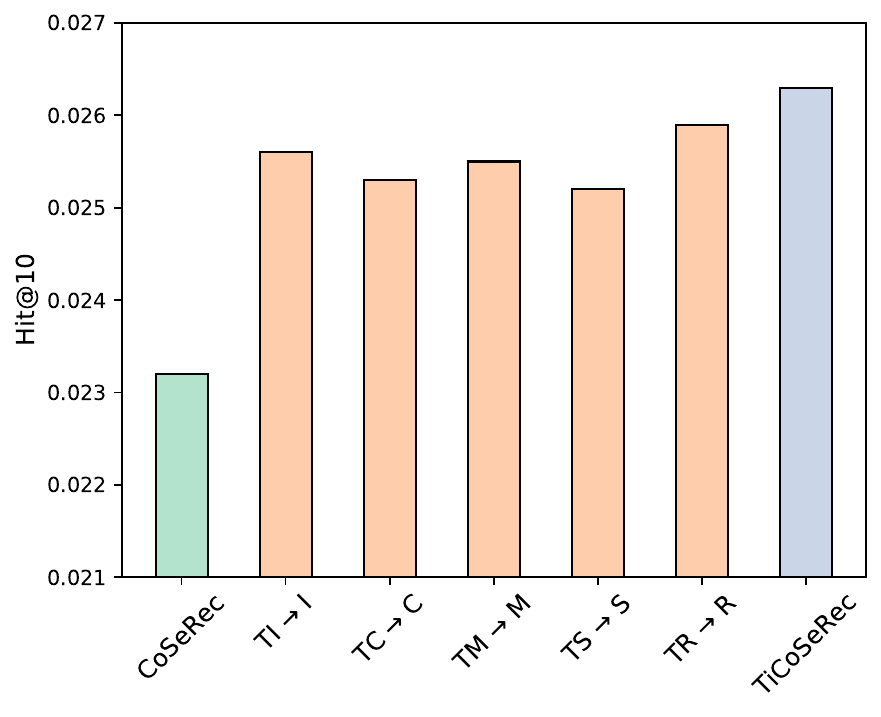}
	}
	\subfigure[Effect of parameter $\sigma$]{
	\includegraphics[scale=0.33]{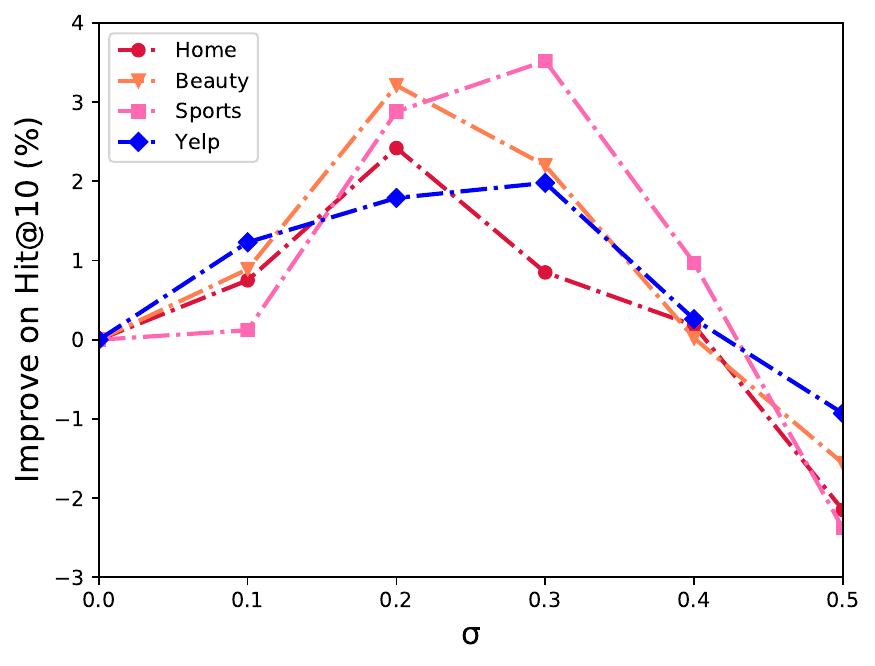}
	}
	\addtocounter{figure}{0}
	\caption{Ablation study on two datasets (Amazon Sports, Home) and sensitivity study on parameter $\sigma$.}
	\label{fig:operator}
	\vspace{-2.5mm}
\end{figure*}

\noindent \textbf{Implementation Details.} We adopt the codes provided by the authors of TiSASRec and CoSeRec, and implement other models in RecBole 1.0 and 2.0 \cite{RecBole, RecBole2}. Note that RecBole uses data augmentation by default. For a fair comparison, we turn off this feature and use our augmented sequence data instead. The training batch size is set to 128. We use the Adam \cite{2014Adam} optimizer with a learning rate of 0.001.

Our approach TiCoSeRec is built upon CoSeRec, which consists of three components. \textbf{(a)} Recommendation by SASRec. \textbf{(b)} Data augmentation by five operators without considering time intervals. \textbf{(c)} Contrastive learning to maximize the agreement between the augmented and original sequences. Hence, the main contributions (and difference) of TiCoSeRec lie in the proposed five data operators with the consideration of time interval.

We set the embedding dimension size as 128 for all models, and take the suggested settings for other parameters as reported in the original papers. For our TiCoSeRec model, we set the number of layers and heads in the Transformer to 2. The maximum sequence length is set to 50. We set parameter $\sigma=0.2$, i.e., top 20\% ranked item sequences are treated as uniform, and no data augmentation will be forced on them. We apply grid search to find the proper values for each data operator's sampling ratio (i.e., $\mu, \eta, \beta$) in the range $[0.1, 0.9]$ (stepping by 0.1) as well as the threshold $K$ to determine short sequences in $\{4,8,12,16,20\}$.

\ \\ \noindent \textbf{Evaluation Metrics.} Following \cite{CoSeRec, 2021CASR}, we adopt the  leave-one-out strategy for performance evaluation. We rank the prediction over the whole item set rather than negative sampling, which otherwise leads to biased discoveries \cite{2020On}. The evaluation metrics include Hit Ratio@k (denoted by HR@k), and Normalized Discounted Cumulative Gain@k (NDCG@k). We report results in HR and NDCG with k $\in \{10, 20\}$. Generally, \emph{greater} values imply \emph{better} ranking accuracy. 

\ \\ \noindent \textbf{Performance Comparison.} Table~\ref{tab:performance} displays the results of all comparison models on four datasets. The column `improve' computes the percentage of improvements obtained by the best approach (ours) relative to the second best one.

For non-sequential models, BPR produces the worst performance, indicating the value of item sequence information. Nevertheless, LightGCN obtains relatively good performance by considering the structural neighborhood information instead. Among all the sequential models without data augmentation, LightSANs achieves the best performance because of the powerful self-attention mechanism in the Transformer structure. Another interesting observation is that TiSASRec works worse than SASRec, though the former takes time interval as additional information for model learning. It implies that time interval has not been well studied in the literature. Our approach opts to use time interval for data augmentation rather than model building. 

Further improvements are obtained by integrating data augmentation with sequential models. It is observed that our approach TiCoSeRec consistently performs the best across all the datasets, and the relative improvements are important and range from 5\% to 18\%. The improvements can be even more significant when compared with CoSeRec, since our main difference is the data augmentation with time intervals rather than traditional operators. Our data augmentation can not only transform non-uniform sequences into uniform ones, but also ensure the newly generated sequences are of high similarity to the original ones.

\ \\ \noindent \textbf{Effect of Data Operators.} Figure~\ref{fig:operator}(a, b) shows the effect of each data operator on overall performance. Specifically, we replace a specific our operator with a corresponding traditional operator (denoted by, e.g., TR$\rightarrow$R) in each time, and fix the other operators unchanged. Although the experiments were conducted on all datasets, results on Yelp and Beauty were omitted due to similar trends and space limitations. 

The recommendation accuracy will be deteriorated when degenerating our data operator to a traditional one, indicating the effectiveness of our data operator by time intervals. Nevertheless, all variants of our approach still outperform the baseline CoSeRec. Hence, it is beneficial for performance improvement to consider all kinds of data operators. Among all the variants, TS$\rightarrow$S has the worst performance while TR$\rightarrow$R reaches the best hit ratio. In other words, the TS operator has the greatest impact on the recommendation performance while the TR operator has the smallest effect. 

\ \\ \noindent \textbf{Effect of Parameter $\sigma$.} This parameter controls the ratio of total item sequences to be treated as uniform, which will not be enhanced by our data operators. The smaller this parameter is, the more item sequences need to be augmented. In our experiments, we tune its value in the range [0.1, 0.5] with stepping 0.1, and we compare its performance improvement with that when all sequences are considered non-uniform (i.e., $\sigma=0$). The results are illustrated in Figure~\ref{fig:operator}(c). 

The results on the four datasets followed similar trends. That is, as $\sigma$ increases, the performance improves gradually and then reaches the maximum value when setting to 0.2 or 0.3. Generally, $\sigma=0.2$ leads to the best performance. In other words, it is the best to consider the top 20\% ranked item sequences as uniform, and the rest needs to be transformed for better accuracy. Further increasing the value of $\sigma$ will not improve performance to a larger extent, but cause much worse accuracy when $\sigma>0.4$. To conclude, a proper setting of parameter $\sigma$ is important for recommendation. 

\section{Conclusions and Future Work}
In this paper, we explored the impact of time interval on sequential recommendations. Our basic idea is that uniform sequences are more valuable for next-item prediction. This assumption was validated by an empirical study. Then, we proposed five data operators to augment item sequences in the light of time intervals. Our experiments on four public datasets have verified the effectiveness of our proposed operators. For future work, we intend to further consider the factor of item category for data augmentation.

\section{Acknowledgement}
This work is partially supported by the National Natural Science Foundation of China under Grant (No. 62032013, 61972078), and the Fundamental Research Funds for the Central Universities under Grant No. N2217004.

{
\bibliography{reference}

\begin{thebibliography}{35}
\providecommand{\natexlab}[1]{#1}

\bibitem[{Du et~al.(2015)Du, Wang, He, Sun, and Song}]{2015Time}
Du, N.; Wang, Y.; He, N.; Sun, J.; and Song, L. 2015.
\newblock Time-sensitive recommendation from recurrent user activities.
\newblock \emph{NIPS}, 28.

\bibitem[{Fan et~al.(2021)Fan, Liu, Lian, Zhao, Xie, and Wen}]{2021LightSANs}
Fan, X.; Liu, Z.; Lian, J.; Zhao, W.~X.; Xie, X.; and Wen, J.-R. 2021.
\newblock Lighter and better: low-rank decomposed self-attention networks for
  next-item recommendation.
\newblock In \emph{SIGIR}, 1733--1737.

\bibitem[{He and McAuley(2016)}]{MC1}
He, R.; and McAuley, J. 2016.
\newblock Fusing similarity models with markov chains for sparse sequential
  recommendation.
\newblock In \emph{ICDM}, 191--200. IEEE.

\bibitem[{He et~al.(2020)He, Deng, Wang, Li, Zhang, and Wang}]{2020LightGCN}
He, X.; Deng, K.; Wang, X.; Li, Y.; Zhang, Y.; and Wang, M. 2020.
\newblock Lightgcn: Simplifying and powering graph convolution network for
  recommendation.
\newblock In \emph{SIGIR}, 639--648.

\bibitem[{Hidasi et~al.(2015{\natexlab{a}})Hidasi, Karatzoglou, Baltrunas, and
  Tikk}]{2015Session}
Hidasi, B.; Karatzoglou, A.; Baltrunas, L.; and Tikk, D. 2015{\natexlab{a}}.
\newblock Session-based recommendations with recurrent neural networks.
\newblock \emph{arXiv preprint arXiv:1511.06939}.

\bibitem[{Hidasi et~al.(2015{\natexlab{b}})Hidasi, Karatzoglou, Baltrunas, and
  Tikk}]{2015GRU}
Hidasi, B.; Karatzoglou, A.; Baltrunas, L.; and Tikk, D. 2015{\natexlab{b}}.
\newblock Session-based recommendations with recurrent neural networks.
\newblock \emph{arXiv preprint arXiv:1511.06939}.

\bibitem[{Jiang et~al.(2021)Jiang, Luo, Kim, Zhang, and Kim}]{2021BiCAT}
Jiang, J.; Luo, Y.; Kim, J.~B.; Zhang, K.; and Kim, S. 2021.
\newblock Sequential Recommendation with Bidirectional Chronological
  Augmentation of Transformer.
\newblock \emph{arXiv preprint arXiv:2112.06460}.

\bibitem[{Kang and McAuley(2018)}]{2018SASRec}
Kang, W.-C.; and McAuley, J. 2018.
\newblock Self-attentive sequential recommendation.
\newblock In \emph{ICDM}, 197--206. IEEE.

\bibitem[{Kingma and Ba(2014)}]{2014Adam}
Kingma, D.~P.; and Ba, J. 2014.
\newblock Adam: A method for stochastic optimization.
\newblock \emph{arXiv preprint arXiv:1412.6980}.

\bibitem[{Krichene and Rendle(2020)}]{2020On}
Krichene, W.; and Rendle, S. 2020.
\newblock On sampled metrics for item recommendation.
\newblock \emph{Communications of the ACM}, 65(7): 75--83.

\bibitem[{Kurashima, Althoff, and Leskovec(2018)}]{2018modeling}
Kurashima, T.; Althoff, T.; and Leskovec, J. 2018.
\newblock Modeling interdependent and periodic real-world action sequences.
\newblock In \emph{WWW}, 803--812.

\bibitem[{Li, Wang, and McAuley(2020)}]{2020Time}
Li, J.; Wang, Y.; and McAuley, J. 2020.
\newblock Time interval aware self-attention for sequential recommendation.
\newblock In \emph{WSDM}, 322--330.

\bibitem[{Li et~al.(2019)Li, Luo, Zhang, Sadiq, and Cui}]{2021Context}
Li, Y.; Luo, Y.; Zhang, Z.; Sadiq, S.; and Cui, P. 2019.
\newblock Context-aware attention-based data augmentation for POI
  recommendation.
\newblock In \emph{ICDEW}, 177--184. IEEE.

\bibitem[{Liu et~al.(2016)Liu, Wu, Wang, Li, and Wang}]{2016context}
Liu, Q.; Wu, S.; Wang, D.; Li, Z.; and Wang, L. 2016.
\newblock Context-aware sequential recommendation.
\newblock In \emph{ICDM}, 1053--1058. IEEE.

\bibitem[{Liu et~al.(2018)Liu, Zeng, Mokhosi, and Zhang}]{2018STAMP}
Liu, Q.; Zeng, Y.; Mokhosi, R.; and Zhang, H. 2018.
\newblock STAMP: short-term attention/memory priority model for session-based
  recommendation.
\newblock In \emph{KDD}, 1831--1839.

\bibitem[{Liu et~al.(2021)Liu, Chen, Li, Yu, McAuley, and Xiong}]{CoSeRec}
Liu, Z.; Chen, Y.; Li, J.; Yu, P.~S.; McAuley, J.; and Xiong, C. 2021.
\newblock Contrastive self-supervised sequential recommendation with robust
  augmentation.
\newblock \emph{arXiv preprint arXiv:2108.06479}.

\bibitem[{Ma et~al.(2020)Ma, Zhou, Yang, Cui, Wang, and Zhu}]{2020Disentangled}
Ma, J.; Zhou, C.; Yang, H.; Cui, P.; Wang, X.; and Zhu, W. 2020.
\newblock Disentangled self-supervision in sequential recommenders.
\newblock In \emph{KDD}, 483--491.

\bibitem[{McAuley, Pandey, and Leskovec(2015)}]{Amazon}
McAuley, J.; Pandey, R.; and Leskovec, J. 2015.
\newblock Inferring networks of substitutable and complementary products.
\newblock In \emph{KDD}, 785--794.

\bibitem[{Qiu et~al.(2022)Qiu, Huang, Yin, and Wang}]{2021DuoRec}
Qiu, R.; Huang, Z.; Yin, H.; and Wang, Z. 2022.
\newblock Contrastive learning for representation degeneration problem in
  sequential recommendation.
\newblock In \emph{WSDM}, 813--823.

\bibitem[{Qu et~al.(2022)Qu, Yuan, Guo, Zhang, and Wei}]{2022cmnrec}
Qu, S.; Yuan, F.; Guo, G.; Zhang, L.; and Wei, W. 2022.
\newblock CmnRec: Sequential Recommendations with Chunk-accelerated Memory
  Network.
\newblock \emph{IEEE Transactions on Knowledge and Data Engineering}.

\bibitem[{Rendle et~al.(2012)Rendle, Freudenthaler, Gantner, and
  Schmidt-Thieme}]{2009BPR}
Rendle, S.; Freudenthaler, C.; Gantner, Z.; and Schmidt-Thieme, L. 2012.
\newblock BPR: Bayesian personalized ranking from implicit feedback.
\newblock \emph{arXiv preprint arXiv:1205.2618}.

\bibitem[{Rendle, Freudenthaler, and Schmidt-Thieme(2010)}]{2010Factorizing}
Rendle, S.; Freudenthaler, C.; and Schmidt-Thieme, L. 2010.
\newblock Factorizing personalized markov chains for next-basket
  recommendation.
\newblock In \emph{WWW}, 811--820.

\bibitem[{Seol, Ko, and Lee(2022)}]{Seol_2022}
Seol, J.~J.; Ko, Y.; and Lee, S.-g. 2022.
\newblock Exploiting Session Information in BERT-based Session-aware Sequential
  Recommendation.
\newblock In \emph{SIGIR}, 2639--2644.

\bibitem[{Sun et~al.(2019)Sun, Liu, Wu, Pei, Lin, Ou, and Jiang}]{2019BERT4Rec}
Sun, F.; Liu, J.; Wu, J.; Pei, C.; Lin, X.; Ou, W.; and Jiang, P. 2019.
\newblock BERT4Rec: Sequential recommendation with bidirectional encoder
  representations from transformer.
\newblock In \emph{CIKM}, 1441--1450.

\bibitem[{Sun et~al.(2020)Sun, Guo, Chen, Zhang, and Wang}]{2020exploiting}
Sun, Y.; Guo, G.; Chen, X.; Zhang, P.; and Wang, X. 2020.
\newblock Exploiting review embedding and user attention for item
  recommendation.
\newblock \emph{Knowledge and Information Systems}, 62(8): 3015--3038.

\bibitem[{Vaswani et~al.(2017)Vaswani, Shazeer, Parmar, Uszkoreit, Jones,
  Gomez, Kaiser, and Polosukhin}]{2017Attention}
Vaswani, A.; Shazeer, N.; Parmar, N.; Uszkoreit, J.; Jones, L.; Gomez, A.~N.;
  Kaiser, {\L}.; and Polosukhin, I. 2017.
\newblock Attention is all you need.
\newblock \emph{NIPS}, 30.

\bibitem[{Wang et~al.(2021)Wang, Zhang, Xu, Chen, Zhang, Zhao, and
  Wen}]{2021CASR}
Wang, Z.; Zhang, J.; Xu, H.; Chen, X.; Zhang, Y.; Zhao, W.~X.; and Wen, J.-R.
  2021.
\newblock Counterfactual data-augmented sequential recommendation.
\newblock In \emph{SIGIR}, 347--356.

\bibitem[{Xie et~al.(2020)Xie, Sun, Liu, Wu, Gao, Ding, and Cui}]{2020CL4SRec}
Xie, X.; Sun, F.; Liu, Z.; Wu, S.; Gao, J.; Ding, B.; and Cui, B. 2020.
\newblock Contrastive learning for sequential recommendation.
\newblock \emph{arXiv preprint arXiv:2010.14395}.

\bibitem[{Xu et~al.(2019)Xu, Ruan, Korpeoglu, Kumar, and Achan}]{time3}
Xu, D.; Ruan, C.; Korpeoglu, E.; Kumar, S.; and Achan, K. 2019.
\newblock Self-attention with functional time representation learning.
\newblock \emph{NIPS}, 32.

\bibitem[{Yao et~al.(2021)Yao, Yi, Cheng, Yu, Chen, Menon, Hong, Chi, Tjoa,
  Kang et~al.}]{2020Self}
Yao, T.; Yi, X.; Cheng, D.~Z.; Yu, F.; Chen, T.; Menon, A.; Hong, L.; Chi,
  E.~H.; Tjoa, S.; Kang, J.; et~al. 2021.
\newblock Self-supervised learning for large-scale item recommendations.
\newblock In \emph{CIKM}, 4321--4330.

\bibitem[{Ye et~al.(2020)Ye, Wang, Chen, Wang, Qin, and Yin}]{time2}
Ye, W.; Wang, S.; Chen, X.; Wang, X.; Qin, Z.; and Yin, D. 2020.
\newblock Time matters: Sequential recommendation with complex temporal
  information.
\newblock In \emph{SIGIR}, 1459--1468.

\bibitem[{Yuan et~al.(2020)Yuan, He, Jiang, Guo, Xiong, Xu, and
  Xiong}]{2020Future}
Yuan, F.; He, X.; Jiang, H.; Guo, G.; Xiong, J.; Xu, Z.; and Xiong, Y. 2020.
\newblock Future data helps training: Modeling future contexts for
  session-based recommendation.
\newblock In \emph{WWW}, 303--313.

\bibitem[{Zhao et~al.(2022)Zhao, Hou, Pan, Yang, Zhang, Lin, Zhang, Bian, Tang,
  Sun et~al.}]{RecBole2}
Zhao, W.~X.; Hou, Y.; Pan, X.; Yang, C.; Zhang, Z.; Lin, Z.; Zhang, J.; Bian,
  S.; Tang, J.; Sun, W.; et~al. 2022.
\newblock RecBole 2.0: Towards a More Up-to-Date Recommendation Library.
\newblock \emph{arXiv preprint arXiv:2206.07351}.

\bibitem[{Zhao et~al.(2021)Zhao, Mu, Hou, Lin, Chen, Pan, Li, Lu, Wang, Tian
  et~al.}]{RecBole}
Zhao, W.~X.; Mu, S.; Hou, Y.; Lin, Z.; Chen, Y.; Pan, X.; Li, K.; Lu, Y.; Wang,
  H.; Tian, C.; et~al. 2021.
\newblock Recbole: Towards a unified, comprehensive and efficient framework for
  recommendation algorithms.
\newblock In \emph{CIKM}, 4653--4664.

\bibitem[{Zhou et~al.(2020)Zhou, Wang, Zhao, Zhu, Wang, Zhang, Wang, and
  Wen}]{2020S3Rec}
Zhou, K.; Wang, H.; Zhao, W.~X.; Zhu, Y.; Wang, S.; Zhang, F.; Wang, Z.; and
  Wen, J.-R. 2020.
\newblock S3-rec: Self-supervised learning for sequential recommendation with
  mutual information maximization.
\newblock In \emph{CIKM}, 1893--1902.

\end{thebibliography}
}

\end{document}